\documentclass{ptptex}

\usepackage{graphicx}

\title{
Galactic modulation of extragalactic cosmic rays:\\
Possible origin of the knee in the cosmic ray spectrum
}

\author{
Hiroshi \textsc{Muraishi}$^{1}$,
Shohei \textsc{Yanagita}$^{2}$
and Tatsuo \textsc{Yoshida}$^{2}$
}

\inst{
$^1$School of Allied Health Sciences, Kitasato University, Kanagawa
228-8555, Japan\\
$^2$Faculty of Sciences, Ibaraki University, Ibaraki 310-8521, Japan
}

\abst{
The existence of the spectral break around $\sim 3 \times 10^{15}$ eV in
the cosmic ray spectrum (referred to as the `knee') is one of the
biggest questions in cosmic ray astrophysics. At the same time, the origin of
cosmic rays above the knee energies (between 10$^{15}$ and 10$^{18}$
eV) is also still unsettled. In this paper, we investigate how the
hypothetical extragalactic CRs after modulated by the galactic wind
contribute to the knee in the CR spectrum. We numerically calculate the
modulated energy spectrum of the hypothetical cosmic rays coming into
the galaxy from just outside of the ``galactic sphere'' where the
galactic wind terminates. We show that the observed knee structure
is reproduced well by a superposition of the modulated component and
the galactic cosmic rays originating in supernova remnants.
}

\begin{document}

\maketitle

\section{Introduction}
Observations have revealed cosmic rays (CRs) spread over 11 decades of
energy from $\sim 10^{9}$ eV to $\sim 3 \times 10^{20}$ eV. The spectrum
is often fitted by power laws of energy with the index of $\sim -2.7$
below $\sim 3 \times 10^{15}$ eV and $\sim -3.0$ above,
respectively. This break in the spectrum is referred to as the `knee' in
the spectrum. The knee has been recognized as an important structure of
the energy spectrum, which may provide constraints on the acceleration
and propagation of CRs.

CRs with energies below the knee have been believed to be originated in
supernova remnants (SNRs) in our Galaxy from general arguments about
energetics and the diffusive particle acceleration mechanism in
shocks.\cite{rf:1,rf:2} \ This argument of the CR origin is
corroborated by recent observations of X-rays (by ASCA) and TeV
gamma-rays (by CANGAROO and HESS) from two SNRs, SN1006
\cite{rf:3,rf:4} \ and
RX\,J1713.7$-$3946.\cite{rf:5,rf:6,rf:7,rf:8,rf:9} \
These observations have revealed that electrons and protons are
accelerated to energies of $\sim$ 100 TeV in the SNRs. \footnote{For
SN1006, there is discrepancy between the results of CANGAROO-I and
HESS; an upper limit reported by HESS is below the flux detected
by CANGAROO-I.\cite{rf:10}} However this maximum energy does not reach
the knee energy of $\sim$ 3 $\times$ 10$^{15}$ eV. A hypothesis blessed
by long tradition explains the break in the spectrum at the knee by more
rapid escape from the Galaxy of more energetic particles above the
knee.\cite{rf:11} \ Alternative interpretations of the knee structure are
to invoke one or a few extra components of galactic origin which dominate
above $\sim$ 1 PeV.\cite{rf:12,rf:13,rf:14,rf:15} \ Yet the origin
of CRs above the knee is still not settled.

Recently the existence of diffuse CR electrons in the intergalactic
space has been suggested from the results of extreme-ultraviolet and
high energy X-ray observation of clusters of galaxies \cite{rf:16} \ 
and from the observation of the diffuse cosmic gamma-ray
background.\cite{rf:17,rf:18} \ If nuclear components with energies
extended
well above $\sim$ 1 PeV also exist together with the diffuse electrons
around our Galaxy, these components modulated by the galactic wind might
be directly observable at the earth. In this paper, we numerically
examine such a possibility, and discuss their implications for the
origin of the knee.

\section{Numerical Simulations}
We postulate the existence of hypothetical CRs just outside of the
``galactic sphere'' where the galactic wind \cite{rf:19} \
terminates. The energy spectrum of these CRs is assumed to be the same
as the spectrum of the CRs observed at the earth with energies higher
than the knee region but extrapolated to much lower energy range; namely
the spectrum is proportional to $E^{-3}$, where $E$ is the total energy
of a particle.

These CRs may diffuse into inner region of the galactic sphere against
the expanding galactic wind. We examine how the spectrum of these CRs
should be modulated during this propagation process. The transport of
CRs is described by the Fokker-Planck equation for the spherical
symetric case \cite{rf:20,rf:21,rf:22} \
\begin{equation}
   \frac{\partial f}{\partial t}
   =\frac{1}{r^{2}}\frac{\partial}{\partial r}
   (r^{2}\kappa\frac{\partial f}{\partial r})
   -V\frac{\partial f}{\partial r}
   +\frac{1}{r^{2}}\frac{\partial}{\partial r}(r^{2}V)
   \frac{p}{3}\frac{\partial f}{\partial p},
\label{eqn:FPE}
\end{equation}
where $f$ is the phase space distribution function, $t$ is the time, $r$
is the radial distance, $V$ is the speed of galactic wind, $p$ is the
particle momentum, and $\kappa$ is the diffusion coefficient for radial
propagation. Here we neglect energy change processes other than
adiabatic losses.

It is known that Eq.(\ref{eqn:FPE}) is equivalent to the coupled
stochastic differential equations (SDEs). The SDEs equivalent to
Eq.(\ref{eqn:FPE}) are written using new quantities $u=\ln(p/mc)$ (where
$m$ is the particle mass and $c$ is the speed of light) as
\begin{equation}
   dr=(V+\frac{2\kappa}{r})dt+\sqrt{2\kappa}\,dW_{r},
   \label{eqn:SDE1}
\end{equation}
and
\begin{equation}
   du=-\frac{2V}{3r}dt,
   \label{eqn:SDE2}
\end{equation}
where $dW_{r}$ is a Wiener process given by the Gaussian distribution,
$P$($dW_{r}$) = ($2\pi dt$)$^{-1/2}$ $\exp$($-dW_{r}^{2}/2dt$). Here we
assume that $V$ does not depend on $r$.
If we integrate the SDEs ``backward in time'',\cite{rf:23,rf:24} \ we can
obtain a probability function $F(p,R|p_{0},r_{0})$ which is necessary to
calculate the modulated energy spectrum at an arbitrary point $r_{0}$
from the Galactic Center. The probability function describes the
probability of which a particle observed with momentum $p_{0}$ at
$r_{0}$ would have had momentum $p$ at the boundary of the galactic
sphere $R$. Once we have calculated $F(p,R|p_{0},r_{0})$, the modulated energy
spectrum $f_{r_{0}}(p_{0})$ at a point $r_{0}$ is calculated with the
energy spectrum $f_{R}(p)$ at $R$ as,
\begin{equation}
f_{r_{0}}(p_{0})=\int f_{R}(p)F(p,R|p_{0},r_{0})dp.
\label{eqn:spectrum}
\end{equation}
This method has been applied successfully to the investigation of the
solar modulation phenomena of the galactic CRs in the
heliosphere.\cite{rf:23,rf:24} \ The momentum spectrum of the
hypothetical CRs
at the boundary $R$ is $f_{R}\propto p^{-5}$, if we assume the spectrum
is proportional to $E^{-3}$ as mentioned earlier.

We integrated numerically Eqs.(\ref{eqn:SDE1}) and (\ref{eqn:SDE2})
assuming the diffusion coefficient $\kappa$ as
\begin{equation}
\kappa = \eta \kappa_{\rm B} \sim 3.3 \times 10^{28} \eta Z^{-1} \!\!\left(\frac{E}{1{\rm PeV}}\right)\!\!\left(\frac{B}{1\mu {\rm G}}\right)^{-1} \rm cm^{2} \rm sec^{-1},
\label{eqn:kappa}
\end{equation}
where $\eta$ is the ratio of the mean free path of the particle to the
Larmor radius and $\kappa_{\rm B}$ is the Bohm diffusion coefficient,
$\kappa_{\rm B}$ = $Ec/(3ZeB)$ (where $E$, $Z$, and $B$ are the total energy
of particle, the atomic number and the magnetic field intensity,
respectively).

The resultant modulated spectrum at an arbitrary
point $r_{0}$ depends on five parameters, $\eta$, $Z$, $R$, $V$, and $B$,
where these parameters are unknown to us. According to the study of the
galactic MHD wind driven by CRs in a rotating galaxy,\cite{rf:25} \ it
is expected that our Galaxy is
surrounded by a large galactic wind halo with a scale of the order of $R
\sim$ 100 kpc. Also, the resultant velocity profile reveals that it
increases almost linearly with distance from the galactic disk up to 20
kpc and then becomes a constant with about $\sim$ 300 km s$^{-1}$ up
to $\sim$ 100 kpc. When the extragalactic CRs propagate into inner
region of the galactic sphere, it will spend long times in the
outer region of the galactic sphere rather than the inner region like
around the galactic disk. Therefore we should select $R \sim$ 100 kpc
and $V \sim$ 300 km s$^{-1}$ in our calculation.

\begin{figure}[top]
\centerline{\includegraphics[width=14cm]{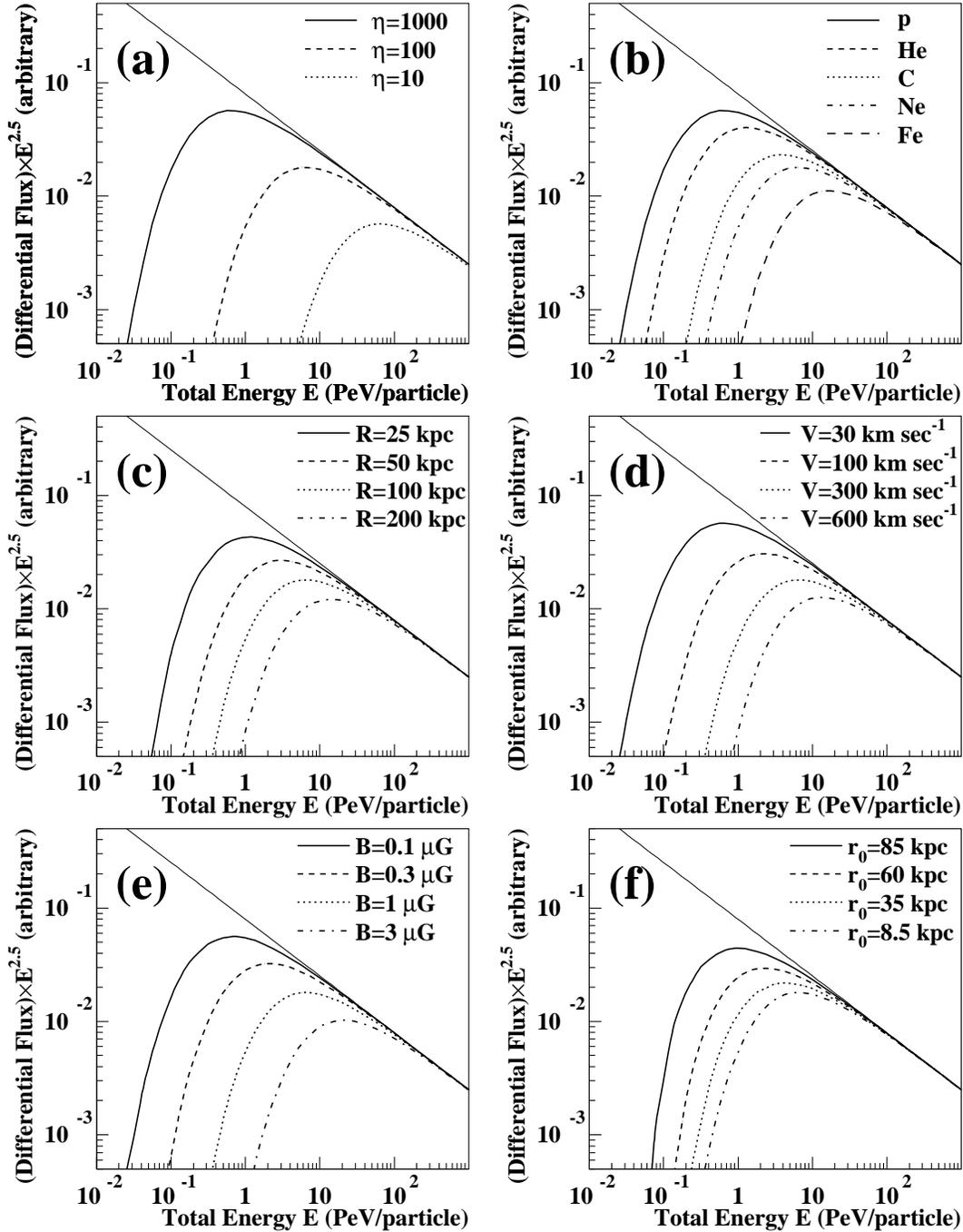}}
\caption{Expected energy spectra of the hypothetical CRs modulated by
 the galactic wind. The straight line indicates the differential energy
 spectrum of the hypothetical CRs with $Z$ at the boundary of the
 galactic sphere $R$ which is a power law in total energy with a
 spectral index of $-$3.0. (a)the differential intensity of protons
 ($Z$=1) at the earth ($r_{0}=8.5$ kpc) for $R=100$ kpc, $V=300$ km
 sec$^{-1}$, and $B=1\mu$G with $\eta$ = 1000(solid line), 100(dashed
 line), and 10(dotted line), respectively. (b)Same as (a) but with
 five values of $Z$ for $\eta$ = 1000, $R=100$ kpc, $V=300$ km
 sec$^{-1}$, and  $B=1\mu$G. (c)Same as (a) but with four values of $R$
 for $\eta$ = 100, $Z$=1, $R=100$ kpc, $V=300$ km sec$^{-1}$, and
 $B=1\mu$G. (d)Same as (a) but with four values of $V$ for $\eta$ =
 100, $Z$=1, $R=100$ kpc, and $B=1\mu$G. (e)Same as (a) but with four
 values of $B$ for $\eta$ = 100, $Z$ =1, $R=100$ kpc, and $V=300$ km
 sec$^{-1}$. (f) Same as (a) but at four values of an arbitrary point
 $r_{0}$ for $\eta$ = 100, $Z$=1, $R=100$ kpc, $V=300$ km sec$^{-1}$,
 and $B=1\mu$G.}
\label{spectrum-p1}
\end{figure}

\section{Results}
\subsection{Galactic Modulated Spectrum}
Fig.~\ref{spectrum-p1}(a) shows the calculated differential energy
spectra of protons ($Z$=1) at the earth ($r_{0}$ = 8.5kpc) as a function
of total energy $E$. The solid, dashed, and dotted lines are the
differential intensity with $\eta$=1000, 100, and 10, respectively. Here
we assume $V=300$ km sec$^{-1}$, $R=100$ kpc, and $B=1\mu$G.
The straight line indicates the assumed unmodulated spectrum at the
boundary of the galactic sphere. The break point of the spectrum should
be compared with the knee. We can find that the spectrum is break around
the knee energy. We also find that if $\eta$ increases
by some factor, the break point is shifted to the lower energy by the
same factor. We also calculated the modulated energy spectra of
hypothetical nuclear components instead of protons. The results are
shown in Fig.~\ref{spectrum-p1}(b), where each lines are the
differential intensity of components labeled in the figure at the earth
with $\eta=1000$ as an example. We find that the break point of the
nuclear components is shifted to the higher energy by a factor of $Z$
compared with that for protons. It may also be worth remarking how is
the break point shifted when the other parameters, $B$, $R$, and $V$ are
changed. The resultant spectra are shown in
Fig.~\ref{spectrum-p1}(c)-(e).
Here if we define the breaking energy $E_{\rm break}$ as the energy in
which the modulated spectrum shown in Fig.~\ref{spectrum-p1} becomes a
maximum, it is approximately shown as
\begin{equation}
E_{{\rm break}}(Z,\eta,B,R,V)\sim
6 \times 10^{15} Z \eta_{\star}^{-1} B_{\star} R_{\star} V_{\star} ~ \rm eV,
\label{eqn:break}
\end{equation}
where $\eta_{\star}$=$\eta$/(100), $B_{\star}$=$B$/(1$\mu$G),
$R_{\star}$=$R$/(100kpc), and $V_{\star}$=$V$/(300 km s$^{-1}$),
respectively. This results indicate clearly that we can reproduce the
knee by choosing an appropriate values for these parameters.

In addition, in order to investigate how the spectrum is modulated in
arbitrary points other than the earth, we calculated the
modulated spectra at four values of $r_{0}$. The results are shown in
Fig.~\ref{spectrum-p1}(f). We found that if $r_{0}$ approaches the
boundary $R$, the break point of the spectrum shifts to low energy more
and the modulated spectrum diminishes more rapidly as energies go down
at the same time. This result indicates that the hypothetical
extragalactic CRs with energy lower than at least TeV cannot penetrate
into the inner region of our galactic sphere due to the effect of
modulation by the galactic wind.

\begin{figure}[top]
\centerline{\includegraphics[width=7.5cm]{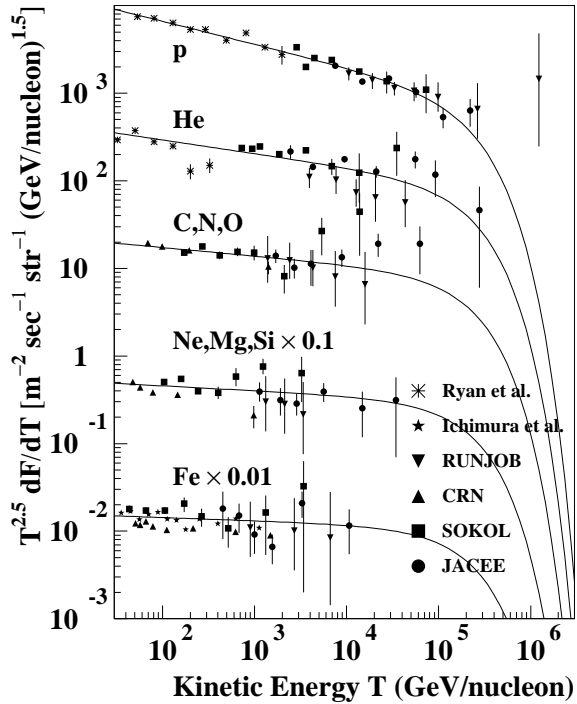}}
\caption{Differential energy spectra obtained by various direct
measurements as a function of kinetic energy per nucleon and the
best fit curves for each element. The data were adopted from
 Refs.~\citen{rf:26,rf:27,rf:28,rf:29,rf:30,rf:31}.}
\label{spectrum-add-n}
\end{figure} 

\subsection{Model Spectrum (Two Components Model)}
In this section, to demonstrate how well our model reproduces the
observed all-particle spectrum near the knee region, we consider a
simple model which is mentioned below. We assume that the all-particle
spectrum observed at the earth $F_{\rm total}(E)$ is superposition of
the two components, namely the modulated extragalactic component $F_{\rm
modul}(E)$ and the component originated in SNRs in our Galaxy $F_{\rm
SNR}(E)$ as
\begin{eqnarray}
F_{\rm total}(E) = F_{\rm SNR}(E) + F_{\rm modul}(E),
\end{eqnarray}
and $F_{\rm SNR}(E)$ and $F_{\rm modul}(E)$ are represented as
\begin{eqnarray}
F_{\rm SNR}(E)&=&\sum_{Z} f_{\rm SNR}^{Z}(E)\\
F_{\rm modul}(E)&=&\sum_{Z} f_{\rm modul}^{Z}(E),
\end{eqnarray}
where $f_{\rm SNR}^{Z}(E)$ and $f_{\rm modul}^{Z}(E)$ are each nuclear
components, respectively.
Now we consider that $F_{\rm SNR}(E)$ consists of the sum of
components after propagated in intersteller space from each of SNRs in
our Galaxy, in which particles are accelerated to a power law by shock
acceleration. Thus we simply assume that each nuclear component
observed at the earth $f_{\rm SNR}^{Z}(E)$ is represented by a power law
with each spectral index of $\alpha$ as
\begin{equation}
f_{\rm SNR}^{Z}(E) \propto E^{-\alpha} {\rm exp} (-E/(ZE_{\rm max})),
\label{eqn:fitting}
\end{equation}
where $E_{\rm max}$ is the maximum energy of protons which may come from
the maximum energy attained by protons accelerated in SNRs. Here we
reexamined published data of CRs which are believed to originate in SNRs
in our Galaxy as shown in Fig.~\ref{spectrum-add-n} as a function of
kinetic energy per nucleon $T$. Here we fit
with $f_{\rm SNR}^{Z}(T)$ $\propto$ $T^{-\alpha}\exp$ ($-T$/$(ZE_{\rm
max}$/$A$)) because the published data is represented as a function of
$T$. We assumed in this fitting $E_{\rm max}$ = 500 TeV for example,
and also assumed that ($Z$, $A$) are (7, 14) for CNO-group and (12, 24)
for NeMgSi-group, respectively. We define the sum of these components
shown in Fig.~\ref{spectrum-add-n} as SNR component $F_{\rm SNR}(E)$.

\begin{figure}[top]
\centerline{\includegraphics[width=14cm]{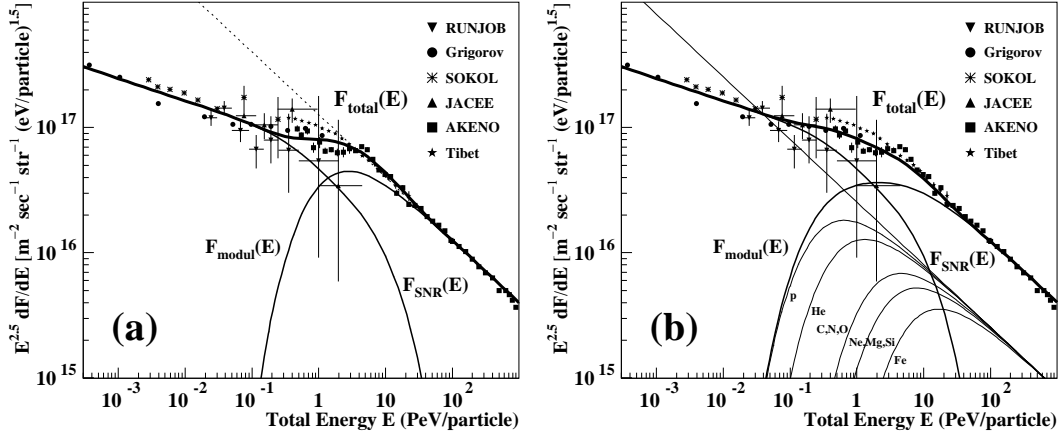}}
\caption{Observed all-particle spectrum and the model spectrum around
the knee versus total energy of particle $E$. The observed data were
adopted from Refs.~\citen{rf:27,rf:30,rf:31,rf:32,rf:33,rf:34}.
The thick solid line indicates the model all-particle spectrum
$F_{\rm total}(E)$ which is the superposition of two components shown by
two solid lines; SNR component $F_{\rm SNR}(E)$ and modulated
extragalactic component $F_{\rm modul}(E)$. The SNR component is the
sum of all nuclear components presented in Fig.~\ref{spectrum-add-n}
but presented here as functions of different energy scale.
(a)Single component model for $F_{\rm modul}(E)$. Modulated
component is calculated for proton $f_{\rm modul}^{\rm p}(E)$ with
$\eta$ = 250 and with the same values for $R$, $V$, and $B$ as in
Fig.~\ref{spectrum-p1}(a).
Straight dotted line indicates the energy spectrum
of the hypothetical CRs at the boundary $R$ as 
$\sim$ 4.3 $\times$ 10$^{20}$ $E^{-3}$ cm$^{-2}$ sec$^{-1}$
str$^{-1}$ eV$^{-1}$.
Notice that we can replace $f_{\rm modul}^{\rm p}(E)$ with
another nuclear component $f_{\rm modul}^{Z}(E)$ (see in the text in
 details).
(b)Composite model for $F_{\rm modul}(E)$. Modulated component consists
 of the sum of five chemical comonents ($f_{\rm modul}^{\rm p}(E)$,
 $f_{\rm modul}^{\rm He}(E)$, $f_{\rm modul}^{\rm CNO}(E)$,
$f_{\rm modul}^{\rm NeMgSi}(E)$, and $f_{\rm modul}^{\rm Fe}(E)$). Each
 components is calculated with $\eta=250$, $R=100$ kpc, $V=300$ km
 sec$^{-1}$, and $B=1\mu$G, where each spectrum assumed at
 the boundary $R$ is the same (shown as straight dotted line) and is
 five times smaller than that for the case in
 Fig.~\ref{spectrum-add}(a). In both figures, the numerical factor in
 $\kappa$ was chosen so that the model spectrum reproduces best the
 observed spectrum.
}
\label{spectrum-add}
\end{figure}

Fig.~\ref{spectrum-add} show the resultant model spectrum
$F_{\rm total}(E)$ around the knee for two example cases with the
all-particle data obtained by various
experiments. Fig.~\ref{spectrum-add}(a) shows a model spectrum when
we adopt $F_{\rm modul}(E)$ for the modulated proton spectrum $f_{\rm
modul}^{\rm p}(E)$ calculated with
$\eta=250$, $R=100$ kpc, $V=300$ km sec$^{-1}$, and $B=1\mu$G. As
clearly seen in Fig.~\ref{spectrum-add}(a), we find our model reproduces
the observed spectrum around the knee fairly well. We emphasize that we
can replace the modulated proton spectrum $f_{\rm modul}^{\rm p}(E)$
with another nuclear component $f_{\rm modul}^{\rm Z}(E)$ calculated
with the values according to the roles shown in Eq.(\ref{eqn:break}).
For example, if we adopt $\eta=500$, $R=100$ kpc, $V=300$ km
sec$^{-1}$, and $B=0.077\mu$G, we can replace $f_{\rm modul}^{\rm p}(E)$
with the modulated iron spectrum $f_{\rm modul}^{\rm Fe}(E)$.

On the other hand, we don't know the chemical composition of the
hypothetical extragalactic CRs. As an extreme example, we assume
compound $F_{\rm modul}(E)$ which consists of the same spectrum
of $f_{\rm modul}^{\rm p}(E)$, $f_{\rm modul}^{\rm He}(E)$, $f_{\rm
modul}^{\rm CNO}(E)$, $f_{\rm modul}^{\rm NeMgSi}(E)$, and $f_{\rm
modul}^{\rm Fe}(E)$ at the boundary. The result is shown in
Fig.~\ref{spectrum-add}(b), where we adopt $\eta=250$, $R=100$ kpc,
$V=300$ km sec$^{-1}$, and $B=1\mu$G, and each abundance ratio at the
boundary is five times smaller than that for the case in
Fig.~\ref{spectrum-add}(a). We can also reproduce a spectrum by using
the model with compound of $F_{\rm modul}(E)$.
We demonstrated our model for a simplistic case of spherical
symmetric geometry with specific forms of $\kappa$.

\subsection{Mean Mass of CRs around the Knee}
\begin{figure}[top]
\centerline{\includegraphics[width=8.3cm]{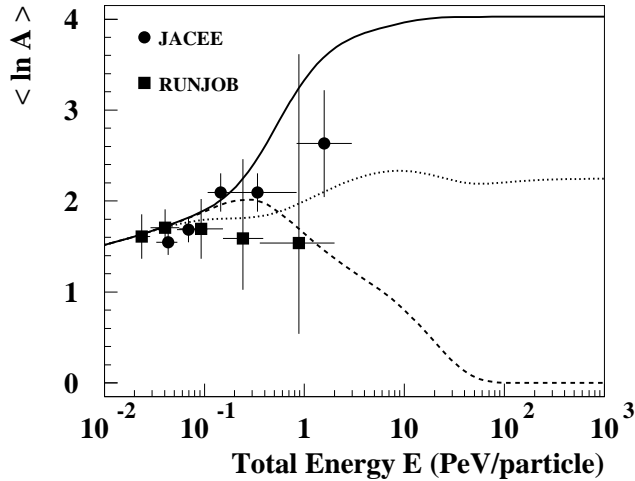}}
\caption{Mean mass of CRs around the knee as a function of total
energy of particle. Symbols indicate the values obtained by direct
measurements.\cite{rf:30,rf:31} \
Dashed line; $F_{\rm total}(E)$ = $F_{\rm SNR}(E)$ +
$f_{\rm modul}^{\rm p}(E)$ (shown in Fig.~\ref{spectrum-add}(a)).
Solid line; $F_{\rm total}(E)$ = $F_{\rm SNR}(E)$ +
$f_{\rm modul}^{\rm Fe}(E)$ (shown in Fig.~\ref{spectrum-add}(a)).
Dotted line; $F_{\rm total}(E)$ = $F_{\rm SNR}(E)$ +
$f_{\rm modul}^{\rm p}(E)$ + $f_{\rm modul}^{\rm He}(E)$ +
$f_{\rm modul}^{\rm CNO}(E)$ + $f_{\rm modul}^{\rm NeMgSi}(E)$ +
$f_{\rm modul}^{\rm Fe}(E)$ (shown in Fig.~\ref{spectrum-add}(b)).
}
\label{dependence}
\end{figure}

On inspection of Fig.~\ref{spectrum-add}, we also find the mean mass
of the CRs around the knee should change with energy in complicated
manner, because SNR component diminishes gradually depending on its
charge and the modulated component creeps in as energies go up.
We have to estimate the expected mean mass to be compared with
measurements by assuming the chemical composition of the hypothetical
CRs, because we do not know it. By using the results of our model
spectrum shown in Fig.~\ref{spectrum-add}, we calculated mean mass of
the CRs. The results are shown in Fig.~\ref{dependence} as a function of
total energy $E$ with the mean mass determined by direct measurements in
regions below the knee. Dashed and solid lines indicate the case
for a model with $F_{\rm modul}(E)$ = $f_{\rm modul}^{\rm p}(E)$ and
with $F_{\rm modul}(E)$ = $f_{\rm modul}^{\rm Fe}(E)$ in
Fig.~\ref{spectrum-add}(a), and dotted line with a composite abundance
for $F_{\rm modul}(E)$ in Fig.~\ref{spectrum-add}(b),
respectively. Experimental determinations of the mean mass
in the region above the knee are, however, scattered in rather wide
range for
various cause depending on the methods in ground-based measurements and
also on the interaction models of hadrons in the data
analysis.\cite{rf:35,rf:36,rf:37} \ Our
model predicts the mean mass above $\sim$ 50 PeV would tend to that of
the hypothetical extragalactic CRs as inferred from
Figs.~\ref{spectrum-add} and \ref{dependence}. Our model should be
testified by future experiments in the energy range much higher than the
knee. 

\section{Discussion}
\subsection{Energetics of the hypothetical CRs}
Heretofore we have simply assumed the existence of the hypothetical
extragalactic CRs without any specification of their origin. We should
address some comments on their origin including energetics. 
As an extreme case, if the hypothetical CRs pervade the intergalactic
space uniformly and the spectrum is extended down to their rest mass
energy as $\sim$ 4.3 $\times$ 10$^{20}$ $E^{-3}$ cm$^{-2}$ sec$^{-1}$
str$^{-1}$ eV$^{-1}$ which we have assumed in Fig.~\ref{spectrum-add}(a)
(straight dotted line), their energy density is estimated to be as high
as $\sim$ 294 eV cm$^{-3}$ when we assume they are protons. The density
parameter corresponding to
this energy density is $\Omega h_{70}^{2}$ $\sim$ 0.057 which is much
larger than that of total baryon in the universe of $\Omega h_{70}^{2}$
$\sim$ 0.020 invoked by the Big Bang Nucleosynthesis.\cite{rf:38} \ 
Therefore it is natural to consider that either the spectrum becomes
harder and/or there is a cut off by other reasons in the energy region
lower than the knee, and/or the hypothetical CRs are confined in local
regions surrounding galaxies as well as our Galaxy and/or cluster of
galaxies. Actually if we assume that the hypothetical CR spectrum
becomes harder with the index of 2 in the energy range lower than
10$^{14.5}$ eV, then the expected energy density is estimated to be as
small as 7.9 $\times$ 10$^{-3}$ eV cm$^{-3}$ without any influence on
the resultant modulated spectrum.
Yet there might remain a possibility that some portion of the density
parameter comes from these invisible hypothetical CRs.

\subsection{Hypothetical CRs in SMC}
Measurements with the EGRET telescope show that the density of CRs
inside the Small Magellanic Cloud (SMC) is several times less than in
our Galaxy.\cite{rf:39} \ This fact indicates simply that GeV CRs
may originate in each galaxy, possibly in SNRs.
This results seem to be incompatible with the existence of the
hypothetical extragalactic CRs which we introduced here.
However if SMC have the galactic wind like our galaxy,
the intensity of hypothetical CRs around GeV region inside SMC should be
suppressed due to the modulation effect of SMC itself.
Actually the galactic wind may have collided with that of SMC, LMC, etc.,
and it may have the complicated structure in which the shock waves are
produced on the boundary. In this paper we do not address its detail.
To investigate the behaviour of the hypothetical CRs
in details, we need further simulation study for complex model
using three-dimensional code including a few galactic sphere (for our
Galaxy, LMC, SMC, etc.).

\subsection{Possible Origin of Hypothetical CRs}
There are some scenarios to support the existence of the hypothetical
extragalactic CRs which we introduced here. V\"olk \& Atoyan
\cite{rf:40} \
discussed the existence of non-thermal
hadronic components in the cluster of galaxy in terms of early
starbursts and magnetic field generation in galaxy clusters. They
suggest that relativistic particles would be confined in galaxy clusters
over times longer than the age of the universe.
Another interesting scenario is that it is possible
to accelerate CR protons up to $\sim$ 10$^{18}$ eV by cosmic shock waves
which are generated by large scale structure formation
\cite{rf:41,rf:42} \ and by mergers between clusters of
galaxies.\cite{rf:43} \ 
Miniati et al.\cite{rf:44} \ have carried out a computational study of
production of CR protons at cosmological shocks
and pointed out that a significant fraction of the
total energy associated with baryons inside a cluster could be
stored in CRs as a consequence of diffusive particle
acceleration at structure formation shocks.
In the future, GeV$-$TeV $\gamma$ ray observation of the violent cluster
merger event shown in the evidence for shock acceleration
\cite{rf:45,rf:46} \ might directly lead to the existence of shock accelerated
extragalactic CRs and might clear the relation between the extragalactic
CRs and the CRs above the knee energy.

\section{Conclusion}
We show that the galactic wind can provide a cutoff in about the right
energy range lower than that around the knee. The all-particle spectrum
of CRs around the knee is explained by the superposition of CRs
originated in SNRs in our Galaxy and the hypothetical extragalactic CRs
modulated by the galactic wind. The CRs above the knee region may be
survivors of the extragalactic CRs in the battle against the outflowing
galactic wind. 

\section*{Acknowledgements}
The authors wish to thank to Prof. T.~Yuda (Tibet), Prof. M.~Nagano
(AKENO) and Dr. M.~Ichimura (RUNJOB) for providing the useful data used
in Fig.~\ref{spectrum-add-n} and Fig.~\ref{spectrum-add}.
One of the authors (S.~Y.) thanks Prof. H.~V\"olk and Prof. F.~Aharonian
for their helpful comments in the early stage of this work.
This work has been supported by the Grant-in-Aid for Scientific Research
 of the Ministry of Education, Science and Culture in Japan (12047204).

\end{document}